# EXPLORING TPUS FOR AI APPLICATIONS

Diego Sanmartín, Vera Prohaska
IE Robotics & AI Club
IE University, Madrid, Spain
{dsanmartin, vprohaska}.ieu2020@student.ie.edu

*Abstract*—Tensor Processing Units (TPUs) are specialized hardware accelerators for deep learning developed by Google. This paper aims to explore TPUs in cloud and edge computing focusing on its applications in AI. We provide an overview of TPUs, their general architecture, specifically their design in relation to neural networks, compilation techniques and supporting frameworks. Furthermore, we provide a comparative analysis of Cloud and Edge TPU performance against other counterpart chip architectures. Our results show that TPUs can provide significant performance improvements in both cloud and edge computing. Additionally, this paper underscores the imperative need for further research in optimization techniques for efficient deployment of AI architectures on the Edge TPU and benchmarking standards for a more robust comparative analysis in edge computing scenarios. The primary motivation behind this push for research is that efficient AI acceleration, facilitated by TPUs, can lead to substantial savings in terms of time, money, and environmental resources.

*Keywords*—**TPU, Google, AI Accelerator, Hardware, HPC**

## I. INTRODUCTION

Google TPUs, or Tensor Processing Units, are specialized hardware accelerators for Machine Learning (ML) workloads. They were first introduced in 2016 as part of Google's efforts to improve the performance and efficiency of its ML systems [1]. TPUs are designed with a custom architecture that is efficient at performing matrix operations and processing large amounts of data. All major Artificial Intelligence (AI) network types strongly rely on matrices for their network computations — whether that be a simple Feed-Forward Network (FFN) or a Deep Neural Network (DNN) [2]. The training and inference costs of large AI models are considered computationally expensive tasks that can become a barrier for low-resource applications. They can also be time-consuming tasks, even on powerful hardware. Additionally, recent trends indicate that scale and performance in generative models are highly correlated, pushing AI models to increase in parameter size [3]. Hence, the need for powerful hardware that can accelerate and lower the cost of running these models is trending. TPUs have been proven to accelerate both training and inference of large AI models. They can perform up to 15–30x faster than contemporary GPUs and CPUs, and achieve much higher energy efficiency, with a 30–80x improvement in one Trillion (Tera) Operations per Second (TOPS) per Watt measure [4]. In addition to their efficiency, TPUs are also highly scalable. Google introduced Multislice technology, enabling AI supercomputing, by connecting up to tens of thousands of TPU chips on the cloud []. In this regard, Google has created large pods of TPUs that can train extensive ML models in a distributed manner, such as workloads required in Natural Language Processing (NLP) or Computer Vision (CV) [6]. As a reference, Google's Cloud TPU v4 can handle one quintillion (Exa) Floating Point Operations Per Second (FLOPS) scale processing [7]. This allowed Google to achieve impressive performance gains and has helped the company develop and use some of the world's largest AI models like PaLM [8] 540B or GLaM [9] with 1.2 trillion parameters.

Google has made its TPU hardware available through its cloud computing platform and external consumers are already taking advantage of the performance gains offered by this hardware accelerator. Midjourney, one of the leading text-to-image AI startups, have been using Cloud TPU v4 to train their state-of-the-art generative model [10].

Additionally, in 2018, Google introduced a series of lighter chips (Edge TPU) available for sale under its subsidiary, the Coral brand. The Micro Edge TPU board, with a size of 64x30mm, is capable of 4 TOPS.

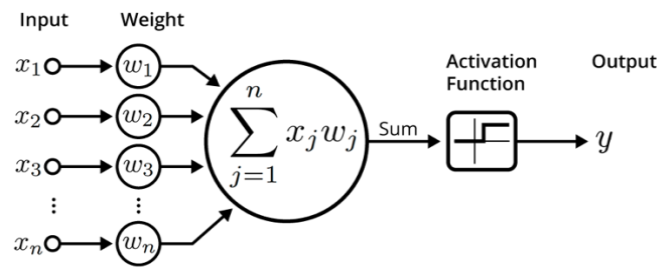

**Fig. 1.** An illustration of an artificial neuron describing the primary operations. [49]

In this paper, we contribute a clear in-depth analysis and definition of the TPUs, their architecture and AI tasks performance metrics for both cloud, and edge computing in comparison to other chip architectures. We further identify the need for the development of new optimization techniques that permit efficient deployment of new AI architectures on the Edge TPU.

## II. LITERATURE REVIEW

Through this literature review, we present an overview of prior studies centered on AI-related TPU research, and comparative studies.

A comparison of the differences between chip architectures, including CPUs, GPUs, FPGAs, ASICs and TPUs, has been studied in [11] focusing on the architectures without any practical experimentation. Other studies, such as [16] and [17] mention TPUs as potential hardware accelerators for parallelization in order to increase AI acceleration. Furthermore, researchers in [18] introduce TPU-MLIR, an end-to-end compiler based on MLIR, designed to deploy pre-trained AI models to TPUs.

Practical implementations often require a benchmark, of which few have been developed for the testing of TPUs. MLPerf [12] is an ML benchmark suite that has gathered the inference and training-speed measures for a large variety of hardware units. The latest version, v3.1, placed the NVIDIA GH200 Grace Hopper Superchip and Google's data center TPU v5e at the top of the benchmark. However, this benchmark only includes the main AI architectures and lacks exploration of the Edge TPUs. [13] compares GPUs against TPUs for accelerating Graph Neural Networks (GNNs), which is an architecture that is not included in the MLPerf benchmark. This demonstrates a lack of completeness in benchmarking comparative research of TPUs.

Deployment of AI architectures on the Edge TPU have been addressed but are also limited. Within the general task of CV, previous research proves superior performance with the implementation of TPUs in running inference on Multiview CNNs [34] combining edge devices (RaspberryPi3 and Edge TPU) for increased FPS [35], and training of Siamese Networks [36]. Authors in [14] explore the performance of a CNN across various edge AI accelerators, and found that the edge TPU outperformed other accelerators in terms of low latency and high accuracy, despite the edge TPU requiring model weights to be quantized to 8bits. [15] compares the performance on four edge platforms but do not include System on Module (SoM) TPU devices. Further CV implementations include pose estimation [38], and semantic segmentation [39, 40].

In terms of NLP, the Edge TPU is more limited. For example, a study done for inference in edge computing of RNNs was not able to conduct experiments on the Edge TPU since the operations from the architecture are not supported [33]. Authors in [19] have managed to deconstruct the encoder Transformer [41] architecture in order to run BERT [42] on an Edge TPU. This limitation similarly applies to generative AI architectures. To the best of our knowledge, there is only applications of GANs that have been fully deployed onto the Edge TPU [37, 50]. With a full deployment we refer to all model layers and operations being compiled by the Edge TPU Compiler [46].

While there is substantial research on TPUs, many of these studies remain isolated and lack a comprehensive exploration or linkage regarding applying TPUs as accelerators in AI development and research.

## III. TPU ARCHITECTURE AND SYSTEM DESIGN

TPUs were built as an Application-Specific Integrated Circuit (ASIC) for Neural Networks (NNs) due to the fast-growing computational demands of AI applications [20]. A classic NN architecture typically involves three primary operations (Figure 1):

1. Multiplication of inputs by corresponding weights.
2. Summation of these products.
3. Application of an activation function to the resulting sum to shape the output as desired.

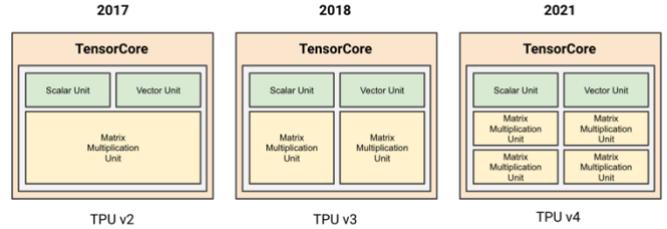

**Fig. 2.** An illustration of the evolution of the Tensor Core. Ranging from 2017 when they introduced the Tensor Core of the TPUv2 with 1 MXU to the TPU v4 Tensor Core which has 4 MXUs and was introduced in 2021.

As can be seen in Figures 2 and 3, TPUs were designed with a Tensor Core composed of specialized hardware components, including Matrix Multiplication Units (MXUs), vector units, and scalar units.

MXUs are optimized for high-speed matrix operations which are fundamental in NNs and represent the most computationally intensive tasks in many ML algorithms. The MXU consists of a 128x128 systolic array of multiply-accumulators, enabling parallelized computations. It delivers the core computational power of the TPU by accelerating the first two primary operations of NNs (matrix multiplication and summation).

Alongside the MXU, the vector unit handles general computations, such as activations and softmax functions. On the other hand, the scalar unit is responsible for tasks like control flow, memory address calculations, and other maintenance operations [21].

Depending on the specific TPU version, whether it's for Google Cloud or the Edge TPU available for purchase, the specifications and capabilities can vary.

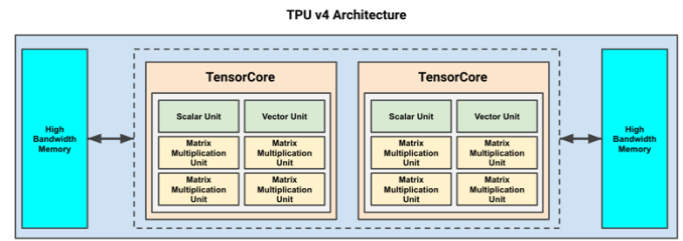

**Fig. 3** Diagram illustrating the architecture of a TPU v4 chip. The architecture contains two Tensor Cores with its corresponding components and a Virtual Core connected to the High Bandwidth Memory.

## A. Datacenter TPUs

TPUs are one of the most powerful and scalable AI hardware accelerators, but they are currently only available for consumers through Google Cloud Platform (GCP). The most recent chip, TPUv5e, provides up to 393 int8 (integral number with 8 bytes depth) TOPS. These can be networked over ultra-fast links to create pods that consist of up to 256 TPUv5e chips and can deliver a total of up to 100 int8 PetaOps of compute power [22]. Pods can also be combined using Multislice, a full-stack almost-linear-performant scaling technology, which connects TPU chips within each slice through high-speed Inter-Chip-Interconnect (ICI) [23]. This allows users to scale to tens of thousands of Cloud TPU chips for individual AI workloads.

## B. The Edge TPU

The Edge TPU series is a set of lighter versions of the datacenter TPUs [24]. For this exploration, we chose to focus only on the System on Module (SoM) devices (Table 1).

While specific internal structures of the Edge TPU remain undisclosed, given that they are an AI ASIC developed by Google's subsidiary (Coral), it very-likely retains the fundamental components of datacenter TPUs such as the MXU, scalar unit, and a vector unit in the chip. Given this, a well-founded estimate of the chip's architecture is an MXU of a 64x64 systolic array with a clock of 480 MHz, resulting in 4 TOPS [21]. It is also worth mentioning the Dev Board Mini integrates a VPU as well as a GPU [25].

Moreover, Edge TPUs exhibit impressive energy efficiency, with the Dev Board Micro processing about 3 TOPS per watt, and the Dev Board Mini and the Dev Board processing at 2 TOPS per watt (refer to Table 1). More interestingly, both the Dev Board Mini and the Dev Board Micro operate without the necessity for cooling. This operational characteristic suggests that buffering likely transpires external to the chip.

| Device | TOPS | TOPS per Watt | Size (mm) | On-chip Processors | SDRAM | DDR |
|---|---|---|---|---|---|---|
| Dev Board | 4 | 2 | 85 x 56 | GPU, VPU, Arm Cortex-A53, Arm Cortex M4 | 1 or 4 GB LPDDR4 SDRAM | 1600 MHz maximum DDR clock |
| Dev Board Mini | 4 | 2 | 64 x 48 | GPU, VPU, Arm Cortex-A35 | 2 GB LPDDR3 SDRAM | 1600 MHz maximum DDR clock |
| Dev Board Micro | 4 | 3 | 65 x 30 | ARM Cortex-M7 and M4 | 64 MB SDRAM | nA |

**Table 1.** High-Level SoM Comparison of the three coral dev boards.

## IV. FRAMEWORKS & COMPILATION FOR TPUS

There are three main frameworks that have been adopted to take advantage of the computational efficiency offered by TPU accelerators: TensorFlow [26], PyTorch [27], and JAX [28]. In the context of this study, our primary emphasis is directed towards the JAX library, since it was designed for XLA compilation and its flexibility for the personalization of AI models.

## A. The XLA Compiler

Code that runs on TPUs must be compiled by the accelerator linear algebra (XLA) compiler. XLA is a domain-specific just-in-time (jit) compiler for linear algebra that takes a graph emitted by an ML framework and compiles its operations into executable binaries [28].

Results illustrated in Figure 4 demonstrate how the application of XLA provides improvements of both speed and memory usage. For example, BERT trained on 8 V100 GPUs using XLA, achieved numbers that approximate a 7x performance in throughput (sequence per second) and a 5x better efficiency in batch size [29]. The reduction in the former (memory usage) enables additional capacity for the implementation of other advanced optimization methodologies, such as gradient accumulation.

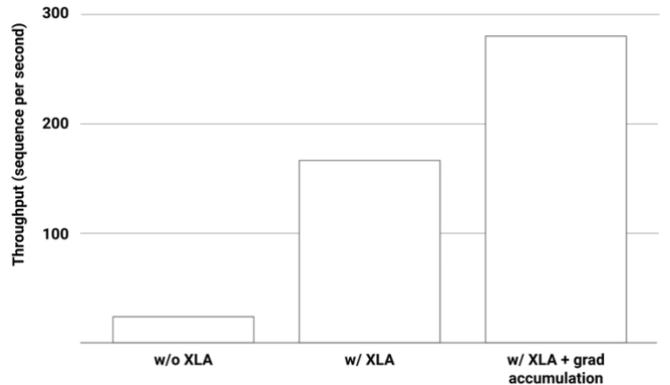

**Fig. 4**. Comparison of training throughput performance on the BERT model using 8 V100 GPUs without (w/o) XLA, with (w/) XLA, and w/ XLA + grad accumulation. [29]

## B. The JAX Library

JAX gives broader flexibility in the development and optimization of AI models since it is a library that was brought together for high-performance ML research [28]. Training a model on GPU/TPU using JAX often provides faster performance and can be cheaper than using PyTorch/XLA on GPU/TPU [10].

It is based on NumPy and takes advantage of XLA to compile and accelerate computations. This is done automatically but can also be defined manually for single Python functions using the @*jit* API decorator. JAX can also automatically differentiate native Python and NumPy functions, which allows for efficient computation of gradients for ML algorithms. It can differentiate between loops, branches, recursion, and closures, as well as advanced derivations of gradients. It supports reverse-mode differentiation (a.k.a. backpropagation) via @*grad* API function as well as forward mode differentiation, and the two can be composed arbitrarily to any order [28].

JAX also has integrations with popular ML libraries such as TensorFlow and PyTorch. In addition, one of the most popular things about JAX is that it is well documented and supported by a large community of researchers.

## V. EXPLORATION OF TPUs

This section explores a practical experimentation of TPUs to measure its performance during inference and training of AI models. Part A summarizes the main differences between the most common chip architectures (CPU, GPU) and the TPU. This serves as an introduction for the subsequent sections which compare the performance of AI workloads in TPUs against CPUs and GPUs.

In Part B, the study explores datacenter TPUs by examining their performance gains in the training of large-scale models.

In Part C, attention shifts to the inference capabilities of the Edge TPU, which holds significant value in the context of Internet of Things (IoT). The motivation behind this study is mainly because running AI models directly on IoT devices minimizes cloud dependency, and makes AI more accessible in our everyday lives.

### A. Comparison OF TPUs, GPUs and CPUs

The most common chip architectures include Central Processing Units (CPUs) and Graphics Processing Units (GPUs). In terms of performance, CPUs and GPUs have different strengths and weaknesses. CPUs are typically better at handling multiple tasks simultaneously. This makes them well-suited to addressing the diverse workloads of a general-purpose computer. On the other hand, GPUs are designed to perform a large number of similar calculations simultaneously, making them ideal for highly parallelizable tasks such as rendering graphics or training model architectures that can be trained in parallel [31]. Unlike CPU, which are general-purpose processors, and GPUs, which were designed to process graphics, TPUs are tailored to the needs of AI algorithms since they leverage matrices as their primary computation instead of only a vector or scalar.

### B. Exploration of Datacenter TPU

In this section we focus on exploring the inference capabilities of cloud-based accelerators, specifically focusing on training a model using CPU, GPU and TPU chips.

The model chosen for the exploration is a faster, lighter version of GPT-2 from OpenAI and has 82 million parameters [32]. The experimentation is conducted using Google Colab [30], which allows for selection of custom runtimes from GCP to accelerate model training. An illustrated comparison of the experiment results can be seen in Figure 5.

For the training, we tokenize and process a dataset to have 17,632 training samples. We then split these samples into batches to end up with a training dataset of 1,102 batches with 16 samples of data.

To test the CPU, we utilized a GCP Intel VM instance with 8vCPUs + 52 GB of memory (n1-highmem-8) and 200GB of SSD. This configuration took 37.07 seconds per iteration which results in 11 hours per training epoch. If we were to do 10 epochs of training, the training time would be at 110 hours or 4 days and 12 hours.

To test the GPU, we attach one NVIDIA T4 GPU to the previous CPU setup. We could already see improvements with this accelerator reducing training to 1.95 seconds per iteration. This would take it 36 minutes per epoch and a total of 6 hours for 10 epochs. These results show that the GPU is 18x faster than the CPU.

To test the TPU, we ran the model training on a TPUv2-8 which provides 8 TPU cores (4 dual-core TPU chips) that allowed us to split the batches of 16 between the 8 cores. Creating batches of 128 samples reduced the total number of batches to be process by each core by $1/8^{th}$. The performance gains of this implementation were significant, taking only 0.68 seconds per iteration, 1 minute and 33 seconds per epoch and a total training time of 13 minutes for the 10 epochs.

These results indicate that a TPUv2-8 VM yields 27x improvement in training over the GPU. Acceleration hardware lowers training from 4 days and 12 hours when employing on a 8vCPU to 6 hours with a T4 GPU to 13 minutes with a Google Colab TPUv2-8 VM.

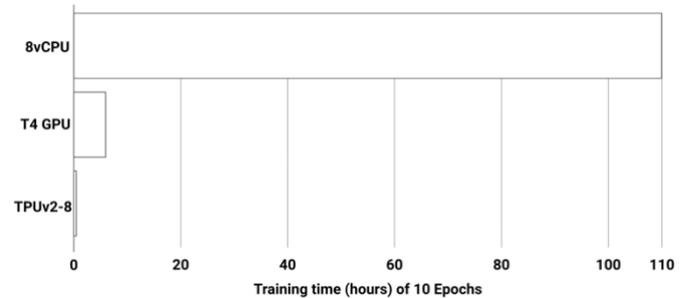

**Fig. 5.** Comparison of training times for 10 epochs between Intel 8vCPU, Nvidia GPU T4, and the TPUv2-8.

### C. Exploration of Edge TPU

In this section we focus on exploring the inference capabilities of smaller SoM devices, specifically the Dev Board Mini TPU, the NVIDIA Jetson Nano, and the RaspberryPi3.

The deployment of AI architectures on the Edge TPU requires the implementation of different optimization techniques [43, 44] to meet the computational overhead and memory requirements of these devices. For one, tensor parameters must be quantized into 8-bit fixed-point numbers, specifically int8 or uint8 formats. Additionally, both tensor sizes and model parameters, including bias tensors, must remain constant during compile time [45]. Furthermore, tensors must be 3-dimensional at most. For tensors exceeding three dimensions, only the innermost three can have a size larger than one. It is also crucial that the model exclusively employs operations compatible with the Edge TPU (see Annex).

Thus, the development of AI models for the Edge TPU, follows a systematic workflow (defined in Figure 6). It starts by ensuring the model aligns with all the supported operations and confirming the constancy of tensor sizes. Finally, the quantized 8-bit model is compiled explicitly for the Edge TPU on a Debian OS, or a Google Colab Notebook [46].

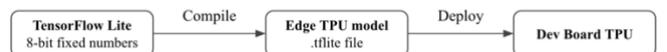

**Fig. 6.** Workflow to compile existing quantized tflite models into Edge TPU executable.

When comparing the devices for speed of inference, we chose to analyze the classification of a 224x224 image using MobileNetV2 [47], provided by the Pycoral library [48].

| Device | Time [s] | Inference FPS |
|---|---|---|
| NVIDIA Jetson Nano (128 CUDA) | 20.52 | 12.18 |
| NVIDIA Jetson Nano (tflite) | 23.02 | 10.86 |
| Raspberry Pi 3 B+ | 59.05 | 4.23 |
| Dev Board Mini Coral TPU | **3.25** | **76.92** |

**Table 2.** Performance of the 3 SoM devices running inference of MobileNetV2 tflite model trained for image sizes of 224x224.

We conducted inference on a single image 250x using MobileNetV2. In order to account for the initialization of the model, and image loading into cache memory, we execute a single prediction first, let the script wait 1 second, and then implement a loop of 250 inference calls of the same image. By measuring the overall duration, we determined the Inference FPS, calculating it as the ratio of processed images to the total inference time. Our results can be found in Table 2.

To provide a fair comparison, we also include the results of the NVIDIA Jetson Nano running the same model using CUDA which is optimized for the NVIDIA GPU. Best results from the Edge TPU and the Jetson indicate a 6x increase in performance when running inference on the TPU, and 18x improvement against the RaspberryPi3 when running inference on MobileNetV2.

## VI. LIMITATIONS AND FUTURE WORK

During the comparison of datacenter machines, due to the limited resource availability of the free tier for Google Colab's TPUs, the exploration of the TPU may not be leveraging its full potential. Executing this on a dedicated TPU VM would allow to find more accurate results. A further limitation is the limited number of operations being supported by the Edge TPU. This may lead many users to prefer the use of Edge GPUs, like the Jetson Nano from NVIDIA, due to the large community support in running models in the GPU.

Furthermore, when comparing edge devices, few benchmarks exist for the direct comparison of device performance. Although some proposals have been made [12, 49], there is little guidance on how to ensure reproducible results for running AI models on edge devices. For example, even running the same model on a GPU vs a TPU can vary, since the TPU requires an 8 bit-quantized model, whereas a Jetson Nano does not explicitly require the quantization of model weights.

In addition to the CPU, GPU and TPU, there are other system-on-chip (SoC) AI accelerator hardware that are pushing the boundaries of ML acceleration. Thus, further investigations could also include the comparison of TPUs against other ML-specific chip architectures. For example VPU, Neural Engine [52], and IBM's AIU [53] are not considered in this exploration, but would be valid candidates for future comparisons.

Finally, in order to compare different architectures, current benchmarks such as MLPerf require an expansion of hardware and models support. The current suite of evaluations also falls short when comparing edge-devices.

## VII. CONCLUSION

In summary, results indicate that the utilization of TPUs in both large datacenter workloads and edge devices can significantly reduce the training and inference times of deep learning models. Their specialized hardware design, tailored for the matrix operations that are most common in neural networks, enables TPUs to achieve accelerated computation speeds and improved power efficiency. We find that through quantization, model compilation and adapting AI architectures to existing ASIC hardware, both training and inference of models can be accelerated exponentially. We leave for future work to investigate further AI architectures and scaling capabilities of TPUs, both in cloud as well as edge computing for the training and inference of AI models, specifically focusing on performance metrics according to a unified benchmark such as MLPerf.

ANNEX

### A. List of supported operations by the Edge TPUs:

| | |
|---|---|
| Add | ReLU |
| Average Pool2d | ReLU6 |
| Concatenation | ReLUN1To1 |
| Conv2d | Reshape |
| Depthwise Conv2d | Resize Bilinear |
| Expand Dims | Resize |
| Fully Connected | Nearest Neighbor |
| L2 Normalization | Rsqrt |
| Logistic | Slice |
| LSTM | Softmax |
| Maximum | Space To Depth |
| MaxPool2d | Split |
| Mean | Squeeze |
| Minimum | Strided Slice |
| Mul | Sub |
| Pack | Sum |
| Pad | Squared-difference |
| PReLU | Tanh |
| Quantize | Transpose |
| ReduceMax | Transpose Conv |
| ReduceMin | |